# Stream Learning: Partition-Fair Gossip Learning Without Tokens

**Fabien Mathieu**
Swapcard, Paris, France
Sorbonne Université, CNRS, LIP6, F-75005 Paris, France

**Alexandre Pham**
Sorbonne Université, CNRS, LIP6, F-75005 Paris, France

**Maria Potop-Butucaru**
Sorbonne Université, CNRS, LIP6, F-75005 Paris, France

**Sébastien Tixeuil**
Sorbonne Université, CNRS, LIP6, F-75005 Paris, France
Institut Universitaire de France, Paris, France

## Abstract

In gossip learning, a network of nodes trains a shared model collaboratively, without a central coordinator, by repeatedly exchanging parts of their local models. The state-of-the-art protocol, Partitioned Token Gossip Learning (PTGL) of Hegedüs et al., splits the weight matrix into $S$ fixed partitions and disseminates them using a token-based fairness mechanism coupled with per-neighbor metadata exchange. We revisit partition scheduling by analogy with peer-to-peer live streaming, where model partitions act as video chunks and partition age acts as chunk scarcity. The analogy yields a design space of two-stage selection strategies (partition first, or neighbor first), from which we instantiate ten concrete protocols collectively called *Stream Learning*. Our main finding is that the simplest of these protocols, which transmits the locally least-trained partition to a uniformly random neighbor (`Ri`), matches PTGL on fault-free workloads while requiring neither token counters nor metadata exchange. Under an adversarial 30% permanent crash of the best-performing nodes, `Ri` matches or outperforms PTGL across all complete-graph configurations tested, with the gap reaching 5.53% on HAR and 5.41% on MNIST in the most heterogeneous regime (Dirichlet $\beta = 0.1$). In our experiments, partition fairness, captured by a single local rule on partition age, accounts for the gap; token-based rate control and utility maximization do not improve over this rule and, under heterogeneity, sit below it.

**Related Version** Conference version to appear in the proceedings of SSS 2026 (Springer LNCS).

**Acknowledgements** Experiments presented in this paper were carried out using the Grid'5000 testbed, supported by a scientific interest group hosted by Inria and including CNRS, RENATER and several Universities as well as other organizations (see https://www.grid5000.fr).

## 1 Introduction

Many machine learning workloads (federated mobile devices, IoT swarms, cross-institution medical data) preclude centralizing the training data on a single server, whether for privacy, regulatory, or bandwidth reasons. Decentralized learning paradigms address this constraint by training a shared model through repeated exchanges of model parameters, with no transfer of raw data.

*Federated Learning* (FL) [14] keeps data local but routes every update through a central server. The server is convenient, but it is also a synchronization bottleneck and a single point of failure. A fully decentralized family of protocols removes it and lets nodes exchange parameters directly with their peers: by gossiping with one neighbor at a time [9, 15], by broadcasting to the whole neighborhood (*epidemic learning* [18]), or through hub-based variants [12]. The resilience of these protocols to faults and adversarial behavior is an active line of work [16]. We work in the *gossip* regime throughout the paper: at each activation, a node sends a representation of its current model to a single neighbor, which merges it into its own.

The design space of partitioned gossip learning is organized around two scheduling questions: (i) *what to send* (the full model, a random sample of parameters, or a fixed partition), and (ii) *when, and to whom.* Hegedüs et al. [10] answer (i) with a partitioning scheme that splits the weight matrix into $S$ disjoint blocks, and (ii) with a **token mechanism** that locally balances proactive periodic sends against reactive sends, in order to prevent both flooding and starvation. The resulting algorithm, Partitioned Token Gossip Learning (PTGL), matches federated learning on the workloads it studies. Subsequent work targets adjacent axes of the same design space rather than the partition-scheduling question itself: Biswas et al. [4] maximize bandwidth use in asynchronous, straggler-prone environments; Biswas et al. [3] synchronize sample exchange in the round-based setting through a per-round gossip matrix; Hu et al. [11] propose a pull-based variant where nodes request samples rather than push them; Biswas et al. [2] add a privacy layer via shared-seed disjoint subsampling.[1] None of these works questions whether the control overhead they add (token counters, synchronization phases, per-neighbor metadata) is actually necessary for partition scheduling.

In this paper, we revisit the *when/who* question by analogy with peer-to-peer live streaming [5]: model partitions play the role of video chunks, and the local age of a partition plays the role of chunk scarcity. The analogy yields a structured design space of two-stage selection strategies. A primary selection picks either the neighbor or the partition first; a secondary selection picks the other under the constraint already imposed. From this space we instantiate ten concrete protocols. The simplest of them, which sends the locally least-trained partition to a uniformly random neighbor (we denote it `Ri`), matches PTGL on fault-free workloads while using neither token counters nor metadata exchange between neighbors. Under 30% permanent node crashes, `Ri` matches or outperforms PTGL across every configuration tested, with the gap widening as data heterogeneity grows. Convergence in partitioned gossip learning appears to be driven by fairness across partitions, not by utility maximization or token-based scheduling; PTGL's token machinery can be dropped without loss of accuracy, with a measurable gain in crash tolerance.

**Contributions.**

- We frame partitioned gossip learning under a peer-to-peer streaming analogy and define an explicit design space of two-stage partition-scheduling strategies (Section 4).
- We instantiate ten concrete strategies, including a lightweight optional metadata channel (*map messages*) for sharing partition ages between neighbors, and evaluate them against PTGL on MNIST and HAR across two Dirichlet-controlled data heterogeneity regimes ($\beta \in \{0.1, 100\}$).
- We study an adversarial crash setting in which the 30% of nodes with the highest individual test accuracy (BoTS) fail permanently, modeling a wipe-out of the most valuable participants.

[1] The fragment construction of Biswas et al. [3, 4] diverges from the model partitioning of Hegedüs et al. [10]. To keep the comparison principled, we adopt the latter throughout this paper.

- The `Ri` strategy matches or outperforms PTGL in every scenario, with the largest margin under adversarial crashes (5.53% on HAR at $\beta = 0.1$ with 30% of the nodes failing). `Ri` uses neither tokens nor the map-message metadata channel: its selection rule reads only the local age vector.

## 2 System Model

Our system is modeled by an undirected, connected graph $G = (V, E)$ with $n = |V|$ nodes. Two nodes $u, v \in V$ can communicate iff $(u, v) \in E$. The state of $u$ is described by its local variables.

A *configuration* of the system is the set of states of all nodes; an *execution* is a sequence of configurations where each transition results from an *event*: transmission of a message between two neighboring nodes, or local computation modifying the state of a node.

We consider a round-based synchronous message-passing system [1]. The execution is divided into rounds. Each node $u$ has an incoming queue $I_u$ and an outgoing queue $O_u$. During a round $r$, each node $u$ can perform a finite number of local actions: processing received messages, periodic activation, local computation. Between round $r$ and round $r+1$, all messages placed in $O_u$ destined for a neighbor $v$ are transferred to $I_v$. Messages are never lost. The communication channels are FIFO and the transmission delays are bounded.

We consider both fault-free and crash-prone models [1]. In the crash-prone model, up to $f < n$ nodes stop executing their code at some point during the execution and never recover.

## 3 Gossip Learning Framework

We briefly recall the gossip learning framework of Hegedüs et al. [10]. Two mechanisms are central to it. A **partition mechanism** reduces message size: instead of transmitting the full set of model parameters, a node sends a single *partition*, a fixed slice of the weight matrix. A **token mechanism** regulates the message flow by interpolating between **proactive** (periodic) and **reactive** (event-triggered) sends, so as to prevent both flooding and starvation. A self-contained background on supervised machine learning and federated learning is provided in Appendix A.

**Partition mechanism.** Each node $u$ holds a fixed local dataset $D_u$ and trains a model $m_u = (w_u, b_u, t_u)$, where $w_u$ are the weights, $b_u$ is the bias, and $t_u \in \mathbb{N}^{S+1}$ is an age vector recording, for each partition $j \in [\![0, S-1]\!]$ and for the bias, the number of updates applied. All nodes share the same architecture and partitioning scheme: a weight matrix $w$ is split into $S$ disjoint partitions $w[0], ..., w[S-1]$. Let $\mathcal{P}(w) \in \{0, ..., S-1\}^{d \times K}$, where $K$ is the number of output classes and $d$ the input dimension, denote the partition assignment of $w$; we use the explicit assignment $\mathcal{P}(w)_{k,l} = (kK + l) \bmod S$, which is one of the round-robin schemes consistent with the partitioning convention of [10]. The bias is shared across all partitions. Eq. (1) illustrates this scheme for $d = 3$, $K = 5$, $S = 3$: each cell of $\mathcal{P}(w)$ holds the partition index of the corresponding weight, and the $K$ entries of $b$ (one per output class, shown as $\bullet$) are shared across all partitions.

$$\mathcal{P}(w) = \begin{bmatrix} 0 & 1 & 2 & 0 & 1 \\ 0 & 1 & 2 & 0 & 1 \\ 0 & 1 & 2 & 0 & 1 \end{bmatrix} \quad b = (\bullet \; \bullet \; \bullet \; \bullet \; \bullet) \tag{1}$$

A node $r$ disseminates a slice of its model via a model message for partition $j$, defined as the 5-tuple $\text{msg} = (j, w_r[j], b_r, t_r[j], t_r[S])$, where $w_r[j]$ denotes the $j$-th partition of $w_r$ (defined above as a slice of weights, indexed by the original coordinates $i$ such that $i \bmod S = j$).

```
m_u = (w_u, b_u, t_u) ← initial values
c_u ← initial values
loop
    Wait(Δ)
    j ← SelectPart()
    with probability σ(c_u[j]) do
        v ← SelectRandomPeer()
        SendModelMessage(v, j)
    else
        c_u[j] ← c_u[j] + 1
procedure OnReceiveModelMessage(msg = (j, w_r[j], b_r, t_r[j], t_r[S]))
    Merge(m_u, msg)
    Train(m_u, D_u)
    x ← φ(c_u[j])
    c_u[j] ← c_u[j] − x
    repeat x times
        v ← SelectRandomPeer()
        SendModelMessage(v, j)
```

**Algorithm 1** Partitioned Token Gossip Learning, running on node $u$. Native asynchronous presentation of [10]; Section 5 details the synchronous mapping used in simulations.

The bias $b_r$ is included in every model message to keep merges consistent. Upon reception, the recipient executes Merge followed by Train.

Both operations modify $m_u$ in place. Train$(m_u, D_u)$ performs one stochastic gradient descent step on $D_u$ (cf. Appendix A) and increments every component of $t_u$ by the batch size. Merge$(m_u, \text{msg})$ destructures the message msg, overwrites $w_{u[j]}$ and $b_u$ with the age-weighted average of the local and received values weighted by $t_{u[j]}, t_{r[j]}$ (resp. $t_{u[S]}, t_{r[S]}$ for the bias), and sets $t_{u[j]}$ and $t_{u[S]}$ to the element-wise maxima of the corresponding local and received ages.

**Token mechanism.** Each node $u$ maintains a counter $c_u \in \mathbb{Z}^S$. On each activation, $u$ samples $j \in [\![0, S-1]\!]$ uniformly; with probability $\sigma(c_u[j])$ (where $\sigma$ is increasing), it sends a *model message* for partition $j$ to a random neighbor, otherwise it increments $c_u[j]$. Upon receiving a *model message* for partition $j$, $u$ merges, trains, then sends $\varphi(c_u[j])$ *model messages* for partition $j$ to random neighbors[2] and decrements $c_u[j]$. The counter acts as a fairness mechanism: a partition that has not been sent for a long time sees its transmission probability increase, while a reception triggers a burst of reactive sends. The full procedure is given in Algorithm 1.

## 4 Stream Learning Framework

We now turn to our contribution, *Stream Learning*. Stream Learning inherits the partition mechanism of PTGL: a *model message* for partition $j$ sent by node $u$ carries the same payload $(j, w_u[j], b_u, t_u[j], t_u[S])$ and triggers Merge then Train on reception. What differs is the rule that decides *which* partition to send *to whom*, and the (optional) metadata used to inform that rule. Following the P2P-streaming analogy of Section 1, we cast partition scheduling as a two-stage selection problem and instantiate an explicit family of strategies. To inform these decisions, Stream Learning introduces a second, lightweight message type: the *map message*, which carries only the age vector $t_u$ ($S+1$ integers, several orders of magnitude smaller than a *model message*). Each node $u$ maintains, for each neighbor $v$, an estimate $\tilde{t}_v$ of $t_v$, updated each time $u$ receives a *map message* from $v$ (we omit the dependence on $u$ for readability). On activation, $u$ sends a *map message* to a random neighbor and a *model message* chosen according to a strategy $s$. We define the *utility* of partition $j$ for a transmission from $u$ to $v$ as $U(u,v,j) = t_u[j] - \tilde{t}_v[j]$, which measures how far $v$ is estimated to lag behind $u$ on $j$. We say $j$ is *useful* for $v$ when $U(u,v,j) \geq 0$. The procedure is given in Algorithm 2.

[2] $\varphi$ takes integer values and is applied on the bounded range of $c_u[j]$, so the number of reactive sends per arrival is bounded.

A strategy $s$ is a two-stage rule: a **primary selection** picks the first dimension (either the neighbor or the partition), and a **secondary selection** picks the second under the constraint imposed by the first. We consider four primary selections (three neighbor-first, one partition-first):

- **R**: choose the neighbor uniformly at random.
- **U**: choose the neighbor $v$ maximizing $|\{j : U(u,v,j) \geq 0\}|$ (the neighbor for whom the most partitions are useful).
- **V**: associate to each neighbor $v$ a partition $j$ drawn uniformly from those maximizing $U(u,v,j)$; pick a neighbor uniformly; let $j$ be its associated partition; finally restrict the neighbor choice to those associated with that $j$ (a vote-like procedure).
- **B** (partition first): choose $j$ maximizing $t_u[j]$ (the most-trained local partition).

And five secondary selections:

- **r**: choose the partition uniformly at random.
- **a**: choose uniformly among partitions of maximal $t_u[j]$.
- **i**: choose uniformly among partitions of minimal $t_u[j]$.
- **m**: if $v$ is fixed, choose $j$ of maximal age among partitions useful to $v$; if $j$ is fixed, choose the neighbor that maximizes $U(u,v,j)$ among the available ones.
- **u**: choose uniformly among partitions of utility $U \geq 0$.

The four primary selections and the five secondary selections yield $4 \times 5 = 20$ combinations. We retain ten of them and discard the others either as semantically redundant (e.g. **B-a** collapses to **B** on its own, while **B-i** contradicts the partition-first primary) or as a priori unpromising (e.g. **U-r** uses utility to pick a neighbor but then makes a uniformly random partition choice, discarding the utility signal on the side where it could most naturally be used). The ten retained strategies are `Rr`, `Ra`, `Rm`, `Ri`, `Ru`, `Bu`, `Vu`, `Vm`, `Uu` and `Um`, summarized in Table 1.

| | **r** | **a** | **i** | **m** | **u** |
|---|---|---|---|---|---|
| **R** | `Rr` | `Ra` | `Ri` | `Rm` | `Ru` |
| **U** | — | — | — | `Um` | `Uu` |
| **V** | — | — | — | `Vm` | `Vu` |
| **B** | — | — | — | — | `Bu` |

**Table 1** The ten Stream Learning strategies retained for evaluation. Rows are primary selections, columns are secondary selections. Cells marked "—" are either semantically redundant with another strategy or a priori unpromising and are not evaluated.

Whenever a selection rule yields multiple candidates with the same maximizer or minimizer, ties are broken uniformly at random throughout the paper.

$m_u = (w_u, b_u, t_u) \leftarrow$ initial values
$\tilde{t} \leftarrow$ initial values

**at each** round **do**
    $v \leftarrow$ SELECTRANDOMPEER()
    SENDMAPMESSAGE($v$)
    $(v, j) \leftarrow$ USESTRATEGY($s, \tilde{t}$)
    SENDMODELMESSAGE($v, j$)

**procedure** ONRECEIVEMAPMESSAGE(msg $= (r, t_r)$)
    $\tilde{t}[r] \leftarrow t_r$

**procedure** ONRECEIVEMODELMESSAGE(msg $= (j, w_r[j], b_r, t_r[j], t_r[S])$)
    MERGE($m_u$, msg)
    TRAIN($m_u, D_u$)

**Algorithm 2** Push-based Stream Learning with strategy $s$, running on node $u$.

**Map messages are optional.** The metadata channel of Algorithm 2 (the $\tilde{t}$ table and the SENDMAPMESSAGE/ONRECEIVEMAPMESSAGE handlers) is only consulted by strategies whose selection rule reads the ages of **partitions held by other nodes**, that is, the **U**-, **V**- and **m**-family strategies. Strategies in the **R**-i / **R**-a / **R**-r / **B**-u families derive their decision from the local age vector $t_u$ alone, in which case the entire map-message machinery can be elided at deployment time. `Ri`, the focus of Section 6, never invokes it.

## 5 Simulation Methodology

**Round-based synchronous execution.** The execution model follows Section 2: time is divided into rounds, each node $u$ has incoming and outgoing queues $I_u$ and $O_u$, and messages placed in $O_u$ at round $r$ are delivered to neighbors' incoming queues by round $r+1$. Within a round, $u$ processes at most one *model message* from $I_u$ (since a merge triggers a local training step, which is the costly operation), but may process all pending *map messages*, which carry only the age vector $t_u$ and require no training. All nodes are activated at the beginning of each round.

**Mapping algorithms to the synchronous model.** PTGL (Algorithm 1) is presented in its native asynchronous form, paced by an activation period $\Delta$; we simulate it by treating each iteration of its activation loop as one round, making $\Delta$ implicit. The proactive draw $\sigma$ remains a per-round Bernoulli, and the reactive burst $\varphi$ is capped at 2 *model messages* per round. With the $\sigma$ and $\varphi$ of Hegedüs et al. [10], the token counters remain bounded and PTGL sends one *model message* per round on average. Stream Learning (Algorithm 2) is presented directly in synchronous form: each node sends one *map message* and one *model message* per round. The incoming queue $I_u$ handles the resulting concurrent arrivals between rounds.

**Gossip and Stream Learning algorithms.** For Partitioned Token Gossip Learning [10] (PTGL), we use the $\sigma$ and $\varphi$ functions of Hegedüs et al. [10], with counters initialized to the average value reported in [7]. We set the number of partitions to $S = 10$. For Stream Learning, the *map message* carries only $S+1$ integers and is several orders of magnitude smaller than a *model message*.

**Datasets and machine learning model.** We use the MNIST and HAR datasets, each split into a training set $\mathcal{D}_{\text{train}}$ and a test set $\mathcal{D}_{\text{test}}$. An MNIST instance is a pair $(x, y) \in \mathbb{R}^{784} \times [\![0, 9]\!]$, where $x$ is a $28 \times 28$ pixel image and $y$ the depicted digit. An HAR instance is a pair $(x, y) \in \mathbb{R}^{561} \times [\![0, 5]\!]$, where $x$ aggregates smartphone sensor readings and $y$ encodes one of six physical activities. Instances of $\mathcal{D}_{\text{train}}$ are partitioned (without replacement) across nodes, so that $D_u \cap D_v = \varnothing$ for $u \neq v$. The characteristics of the datasets are summarized in Table 2.

| Datasets | MNIST | HAR |
|---|---|---|
| Training set size | 60000 | 7352 |
| Test set size | 10000 | 2497 |
| Input dimensions | 784 | 561 |
| Number of classes | 10 | 6 |
| Class imbalance | none | slight |

**Table 2** Characteristics of datasets studied in this paper

Each node $u$ trains a multinomial logistic regression, as in [10] and recalled in Section 3.

**Local datasets.** We control the heterogeneity of the per-node local datasets via a Dirichlet distribution of concentration parameter $\beta$. A high $\beta$ produces near-uniform per-node label distributions (we refer to $\beta = 100$ as the i.i.d. scenario); a low $\beta$ produces a highly skewed split (we refer to $\beta = 0.1$ as the most heterogeneous non-i.i.d. scenario). We illustrate both regimes in Figure 1.

**Topology and scale.** Simulations use a complete graph or a random 20-regular graph, both with $n = 100$. The choice of a complete graph isolates the effect of the selection strategy from the influence of the topological structure, a known confounder in gossip-protocol evaluation. The choice of a random 20-regular graph is motivated by the fact that PTGL's original authors studied their algorithm with a 20-out graph, and that some of our algorithms (e.g., `Vm`) require bidirectional links to work (otherwise, state messages are useless). The choice of $n = 100$ is dictated by the heterogeneous regime: generating a usable Dirichlet split at $\beta = 0.1$ becomes impractical at the larger scales ($n > 4000$) used in earlier gossip-learning studies.[3]

**Metric.** We evaluate both gossip learning and stream learning in terms of average *accuracy*, defined as the fraction of correct predictions on the global test set, averaged over all participating nodes. Higher values indicate better performance.

**Reproducibility.** Each configuration is averaged over five independent runs (15 runs in Section 6.3). The standard deviation across runs is small relative to the curve separations of interest; we therefore omit error bars from the figures for readability. The code is based on gossipy [17] and is available at [13].

**Model variants.** On top of the base protocol of Section 4 and PTGL, we study two variants:

- *Node failures (BoTS).* At a given round, the 30% of nodes with the highest accuracy on the global test set crash simultaneously and permanently. This Best-on-Test-Set policy models an adversarial wipe-out of the most valuable participants. Crashed nodes are excluded from the reported metrics. The failure round is chosen so that the system has had enough rounds to start converging but is not yet at the plateau: $r = 21$ at $\beta = 100$ on both datasets, $r = 16$ at $\beta = 0.1$ on HAR, and $r = 11$ at $\beta = 0.1$ on MNIST (the MNIST runs converge faster, so the failure is triggered earlier on that dataset).

[3] At larger scales, the $\beta = 0.1$ regime produces nodes whose local sets are so skewed that some classes are entirely missing from many local sets, which prevents meaningful local training. We accept the smaller $n$ in exchange for being able to study the regime where the strategies actually differ.

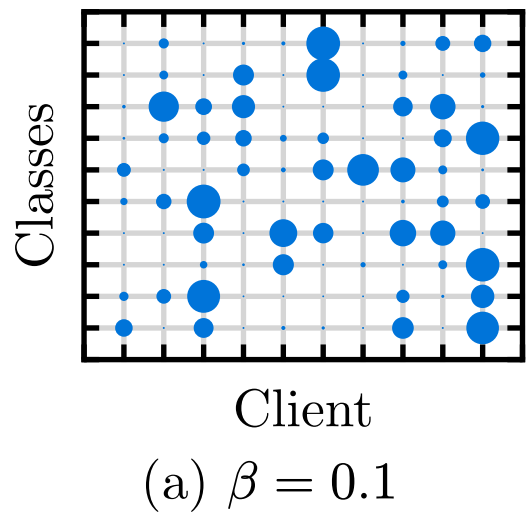

(a) $\beta = 0.1$

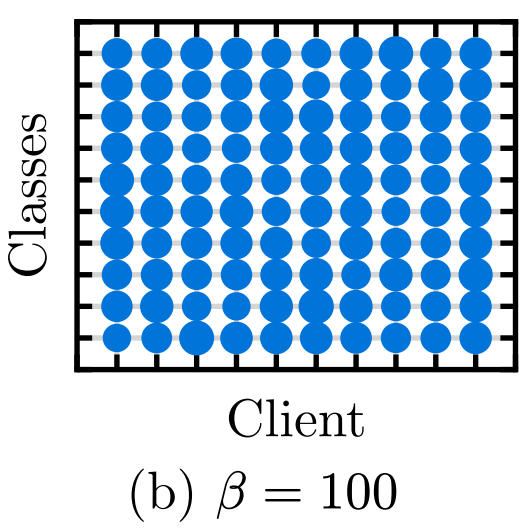

(b) $\beta = 100$

**Figure 1** MNIST train set split across 10 clients using a Dirichlet distribution of parameter varying from $\beta = 0.1$ (left) to $\beta = 100$ (right), shown with 10 clients for visual clarity (experiments in Section 6 use $n = 100$). The larger the circle is for a node and a class, the more data of this class is concentrated in this node.

- *Batched merge.* Nodes process all *model messages* in $I_u$ within a single round: they perform one merge per partition represented in $I_u$ (i.e. from 1 to $S$ merges) before triggering a single local update.

## 6 Simulation Results

We now report our simulation results, under the methodology of Section 5. We proceed by elimination: we start in Section 6.1 by comparing four representatives of the design space of Table 1 against PTGL on the base protocol; we eliminate the strategies that exhibit anti-patterns (explained later in this section). The survivors are then evaluated under node crashes (Section 6.2), and finally under a *batched-merge* protocol variant (Section 6.3).

Of the ten strategies of Table 1, four exhibit qualitatively distinct trajectories and are tracked throughout: `Ri` (the partition-age minimizer), `Ra` (its opposite, the partition-age maximizer), `Rr` (the degree-zero baseline that uses neither tokens nor metadata), and `Vm` (the best representative of the utility-driven family). The remaining six strategies sit between these representatives and are omitted for readability.

Specialized to any **R**-secondary strategy (`Ri`, `Ra`, or the degree-zero baseline `Rr`), the generic protocol of Algorithm 2 collapses to Algorithm 3: no map-message exchange, no token counters, no per-neighbor metadata. The scheduling logic reduces to a single partition-selection rule on the local age vector $t_u$, parameterized by the secondary selector $s$ (cf. Table 1).

### 6.1 Base Scenario

Figure 2 shows accuracy under the fault-free protocol for the four representative strategies `Ri`, `Ra`, `Rr`, `Vm`, and the PTGL baseline `PT`, across both datasets and both heterogeneity regimes in a complete graph topology.

In the homogeneous regime ($\beta = 100$, top row), `Ri`, `Rr` and `PT` converge together to within visible noise on both datasets, plateauing around 0.88. `Ra` and `Vm` both trail by a small margin. With i.i.d. local data, partition-level scheduling is largely redundant: the degree-zero baseline `Rr`, which uses no age and no metadata, tracks the state-of-the-art `PT`.

In the heterogeneous regime ($\beta = 0.1$, bottom row), `Ri` retains the convergence shape it had at $\beta = 100$ and a comparable terminal plateau (within 4 points of accuracy), whereas both `PT` and `Rr` drop visibly. The final ordering is `Ri` $>$ `Rr` $>$ `PT` on both datasets, with `Ri` above `PT` by 14.38% on HAR and 12.15% on MNIST. The ordering reveals two points worth highlighting. First, `PT` plateaus **below** `Rr` by 3.94% on HAR and 7.81% on MNIST: when

$m_u = (w_u, b_u, t_u) \leftarrow$ initial values
**at each** round **do**
  $v \leftarrow$ SELECTRANDOMPEER()
  $j \leftarrow$ PickPartition$(t_u, s)$ ▷ $s \in \{i, a, r\}$
  SENDMODELMESSAGE$(v, j)$
**procedure** PICKPARTITION$(t_u, s)$
  **if** $s = i$ : **return** $\text{argmin}_{k \in \{0,\dots,S-1\}} t_u[k]$ ▷ Ri
  **if** $s = a$ : **return** $\text{argmax}_{k \in \{0,\dots,S-1\}} t_u[k]$ ▷ Ra
  **if** $s = r$ : **return** Uniform$(\{0, \dots, S-1\})$ ▷ Rr
**procedure** ONRECEIVEMODELMESSAGE$(\text{msg} = (j, w_r[j], b_r, t_r[j], t_r[S]))$
  MERGE$(m_u, \text{msg})$
  TRAIN$(m_u, D_u)$

**Algorithm 3** Stream Learning under any **R**-secondary strategy ($s \in \{i, a, r\}$), running on node $u$. Compared to Algorithm 2, the map-message machinery is removed and the partition selection reads only the local age vector. The three secondary rules differ only in how they pick a partition: the least-trained (Ri), the most-trained (Ra), or a uniformly random one (Rr).

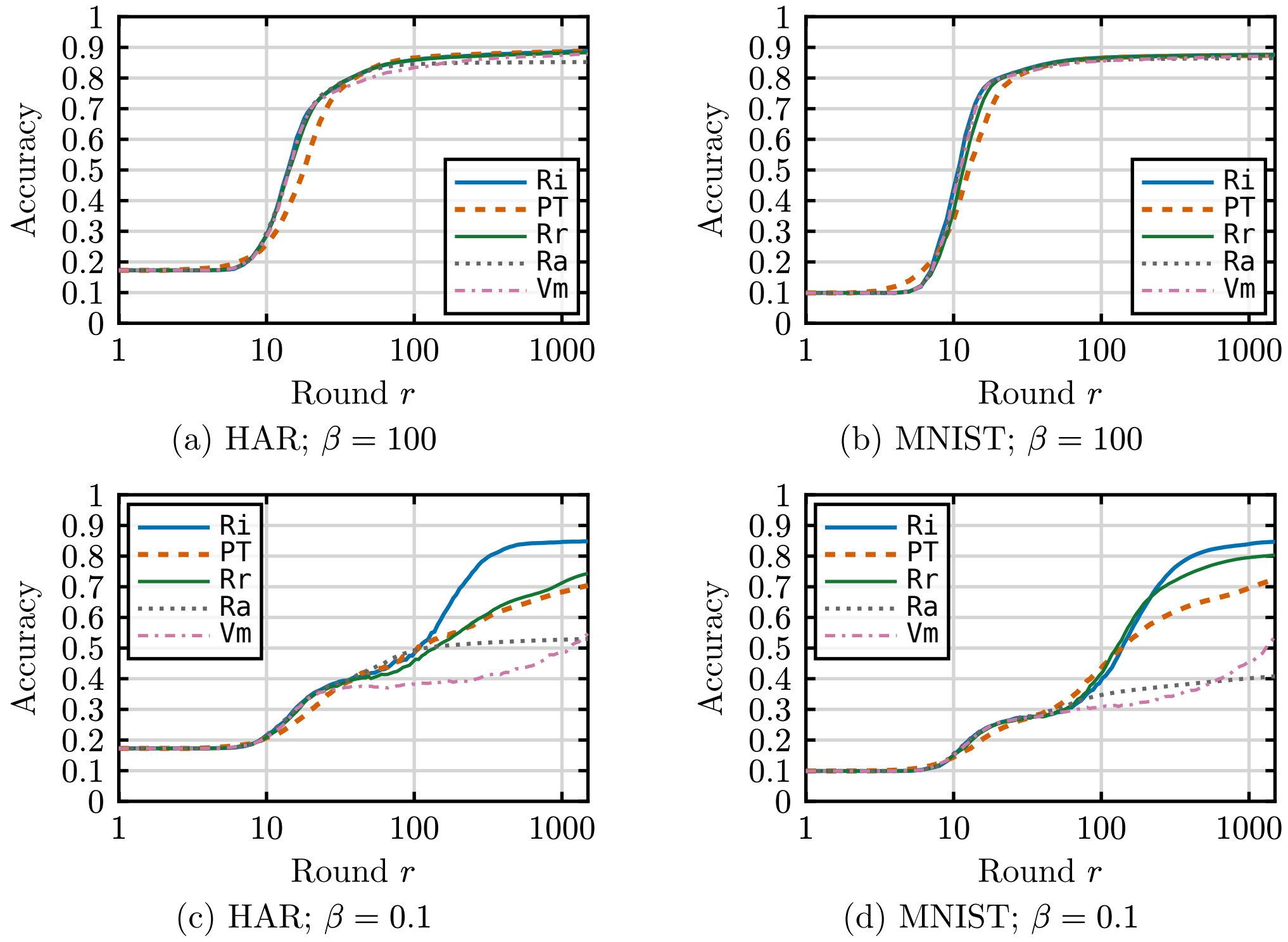

**Figure 2** Complete graph, average accuracy as a function of rounds under the fault-free protocol. Rows correspond to $\beta \in \{100, 0.1\}$, columns to HAR and MNIST.

local data is skewed, the token machinery underperforms uniform random selection, which suggests that its fairness mechanism is misaligned with the partition-level imbalance induced by heterogeneity. Second, Ri leads Rr by 10.44% on HAR and 4.34% on MNIST. This isolates the contribution of the local age vector: that single piece of state, on its own, closes the heterogeneity gap. Neither observation makes use of PT's tokens or of the metadata channel of Algorithm 2; Ri reads only $t_u$, as shown in Algorithm 3.

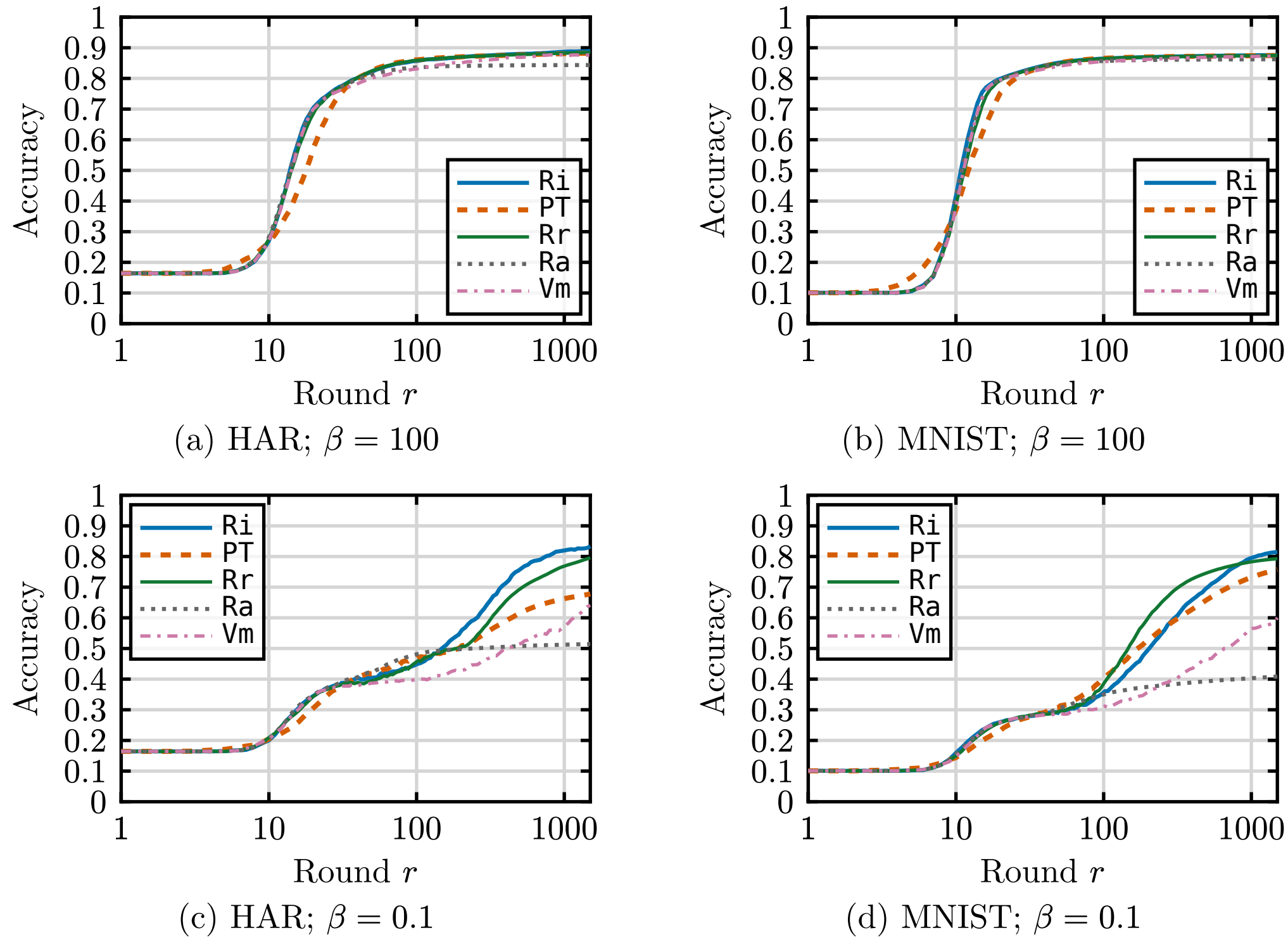

**Figure 3** Random 20-regular graph, average accuracy as a function of rounds under the fault-free protocol. Rows correspond to $\beta \in \{100, 0.1\}$, columns to HAR and MNIST.

**Anti-patterns.** Two strategies exhibit pathological behavior and are eliminated from the rest of the analysis. `Ra`, which transmits the locally **most**-trained partition, sustains a reinforcement loop: high-age partitions get repeatedly sent and trained while others stagnate. The effect is mild at $\beta = 100$ but severe at $\beta = 0.1$ on HAR (`Ra` ends 31.81% below `Ri`). `Vm`, which optimizes utility through map-message-driven coordination, suffers from a related issue already visible at $\beta = 100$: its joint maximization on neighbor and partition tends to send the same partition to the same node repeatedly, producing pseudo-collisions that defeat the intended diversification. At $\beta = 0.1$ the effect compounds with heterogeneity and `Vm` plateaus around 0.54 on both datasets, i.e. 30.15% below `Ri` on HAR and 30.68% on MNIST.

The three surviving strategies `Ri`, `Rr`, and `PT` are carried into the next two subsections, which test whether their ranking is preserved when 30% of the best-performing nodes are removed (Section 6.2) and when nodes batch their merges per round (Section 6.3).

Similarly to Figure 2, Figure 3 shows results under the fault-free protocol, but in a random 20-regular graph. At convergence time, the ranking observed in the complete topology case remains the same here for both homogeneous and heterogeneous distributions. However, we do observe differences in convergence rates and values, especially for `Ri` in the heterogeneous regime ($\beta = 0.1$) for MNIST.

## 6.2 Node failures

Figure 4 overlays the BoTS failure scenario (solid) against the no-failure baseline (dashed) for the three survivors of Section 6.1. The 30% highest-accuracy nodes are removed at the round marked by the vertical green line, then excluded from the accuracy metric, and the run continues in a complete graph topology.

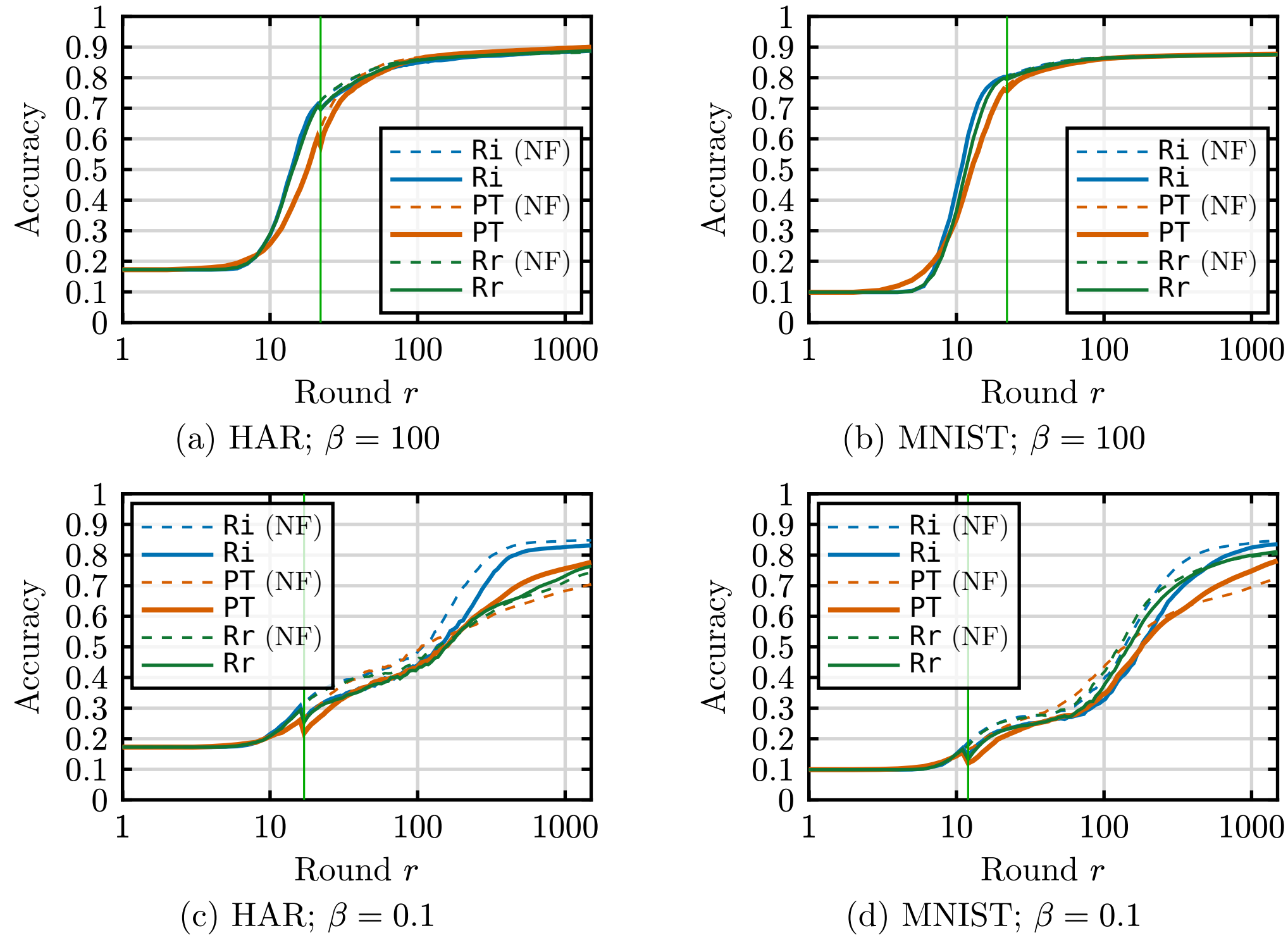

(a) HAR; $\beta = 100$ (b) MNIST; $\beta = 100$ (c) HAR; $\beta = 0.1$ (d) MNIST; $\beta = 0.1$

**Figure 4** Complete graph, average accuracy under best-on-test-set node failures (BoTS). The 30% of nodes with the highest individual test accuracy crash at the round marked by the vertical green line: round 21 at $\beta = 100$ on both datasets (top row), round 16 on HAR $\beta = 0.1$ (panel (c)), and round 11 on MNIST $\beta = 0.1$ (panel (d), as MNIST converges faster). Solid curves are post-crash trajectories (BoTS); dashed curves are the corresponding no-failure baselines (NF) for reference. Rows correspond to $\beta \in \{100, 0.1\}$, columns to HAR and MNIST.

At $\beta = 100$ (top row), BoTS has essentially no effect. `Ri`, `Rr` and `PT` all complete the simulation within visible noise of their no-failure baselines; both `PT` and `Rr` finish very slightly above their NF curves on HAR, which foreshadows the recovery effect we describe below. Removing the 30% best-accuracy nodes does not stress the system in this regime, because the surviving nodes hold equivalent, redundant local datasets.

At $\beta = 0.1$ (bottom row), `Ri` remains clearly ahead under BoTS, but the gaps compress. `Ri` takes a mild post-crash dip (under 2 points of accuracy) and recovers smoothly, ending 5.53% above `PT` on HAR and 5.41% above on MNIST. The `Ri` / `Rr` gap also shrinks but stays in favor of `Ri` (6.74% on HAR, 2.42% on MNIST), while `PT` and `Rr` end close to each other (`PT` slightly ahead on HAR, `Rr` ahead on MNIST). The qualitative conclusion of Figure 2 carries over to the most stressful crash scenario we tested: local age information remains the differentiator.

On the heterogeneous panels (c) and (d), both `PT` and, to a lesser extent, `Rr` eventually rise above their own no-failure baselines under BoTS: `PT` by 7.2 and 5.6 percentage points, `Rr` by 2.05 and 0.78 (HAR and MNIST respectively). **Removing** the 30% best-accuracy nodes thus improves the surviving nodes' average accuracy under these two strategies. `Ri` shows no such effect and tracks its NF baseline closely, slightly below it. We report this pattern without a confirmed mechanistic explanation: it is consistent across our two datasets but rests on a single failure mode (BoTS) and a single state-of-the-art baseline (PTGL), and a confirmation would require instrumenting per-node training trajectories across the failure event, which we leave to future work. A possible explanation is that even in a complete graph, due to the skewed data distribution, gossip learning algorithms are biased as pointed out by [8]. In our cases, nodes that perform well likely create biases towards their own data. Removing those nodes prevents them from continuously biasing the system and hurting overall performance. In essence, `PT` is similar to `Rr` (partitions sent proactively are chosen randomly), which would explain their similar behavior. It would also explain why `Ri` (which always sends the less trained partition, so the partition less likely to be biased) does not show this behavior.

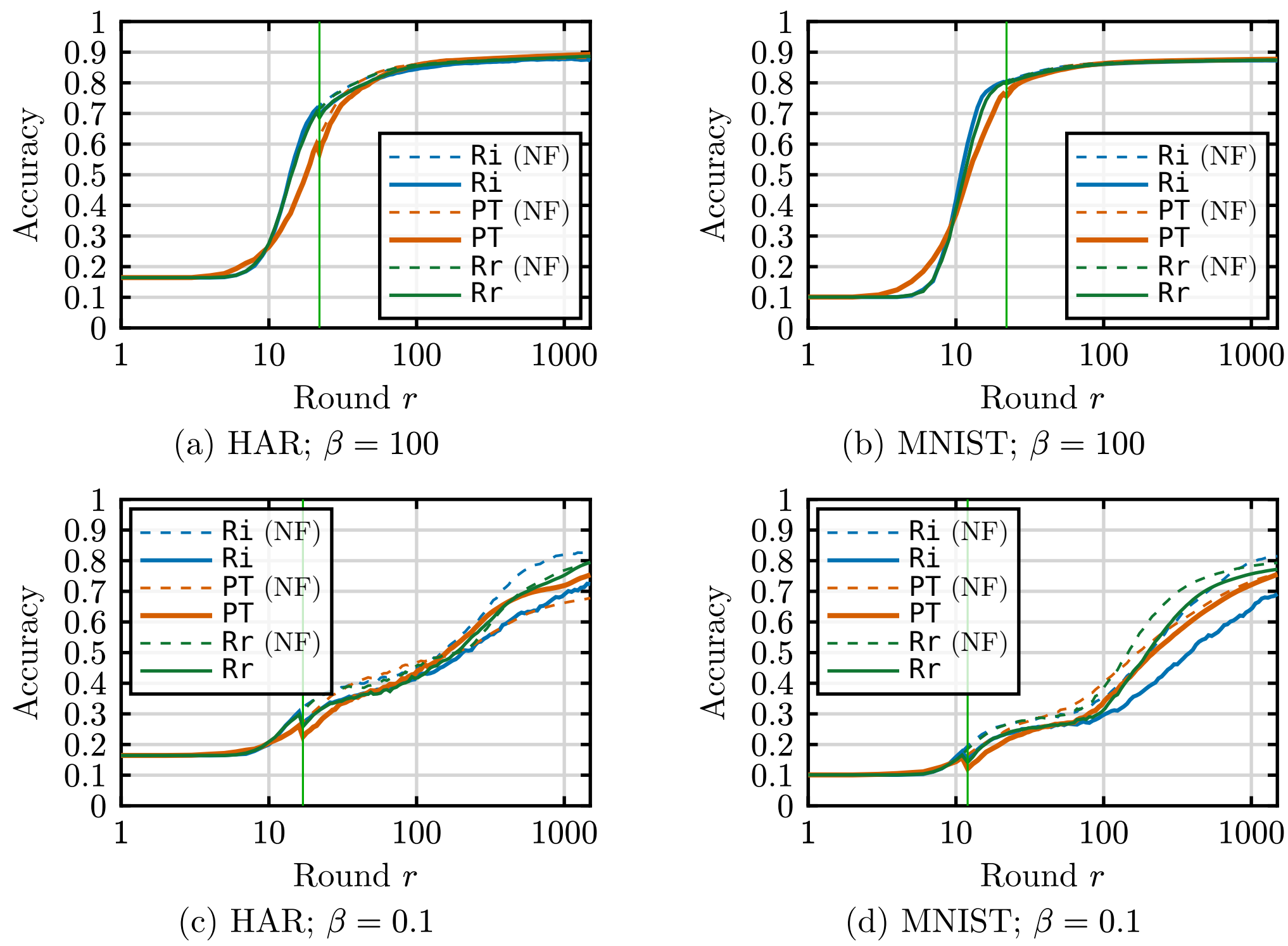

(a) HAR; $\beta = 100$

(b) MNIST; $\beta = 100$

(c) HAR; $\beta = 0.1$

(d) MNIST; $\beta = 0.1$

**Figure 5** Random 20-regular graph, average accuracy under best-on-test-set node failures (BoTS). The 30% of nodes with the highest individual test accuracy crash at the round marked by the vertical green line: round 21 at $\beta = 100$ on both datasets (top row), round 16 on HAR $\beta = 0.1$ (panel (c)), and round 11 on MNIST $\beta = 0.1$ (panel (d), as MNIST converges faster). Solid curves are post-crash trajectories (BoTS); dashed curves are the corresponding no-failure baselines (NF) for reference. Rows correspond to $\beta \in \{100, 0.1\}$, columns to HAR and MNIST.

Similarly to Figure 4, Figure 5 shows results in the BoTS failures scenario, in a random 20-regular graph. Although there is no major difference from the complete graph topology in the homogeneous scenario ($\beta = 100$), the ranking observed earlier in the heterogeneous case is flipped. Under BoTS failures, we observe that `Ri` is slower than both `Rr` and `PT`. Due to the sparse nature of this topology, the fact that `PT` outperforms `Ri` with a 6.38% improvement (resp. 3.02%) on the MNIST (resp. HAR) dataset, while `Rr` outperforms `Ri` with a 8.03% improvement (resp. 7.04%) on the MNIST (resp. HAR) dataset suggests that under BoTS failures and specific topology, choosing a partition randomly is the decisive factor to recover faster from the failures, as `Ri` is inefficient here.

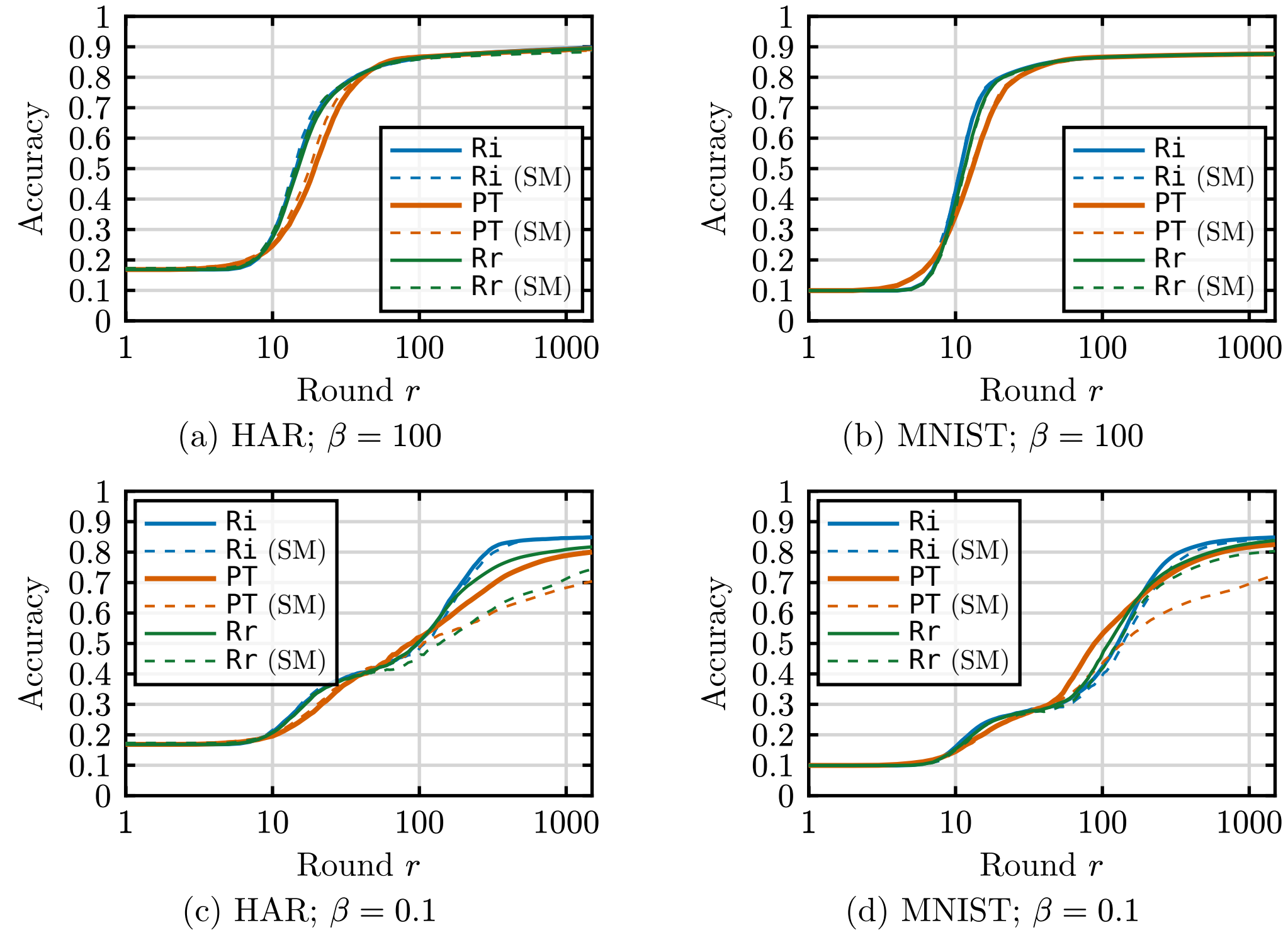

**Figure 6** Complete graph, batched-merge (solid) vs single-merge (dashed) for `Ri`, `Rr`, and `PT`. Rows: $\beta \in \{100, 0.1\}$. Columns: HAR, MNIST. At $\beta = 100$ (top row) BM and SM are visually indistinguishable for all three strategies. At $\beta = 0.1$ (bottom row), BM has essentially no effect on `Ri` (already near-saturation) but lifts both laggards: roughly 10 points of accuracy on `PT` and 3 to 7 on `Rr`.

Also on the heterogeneous case, we still observe (to a lesser degree) the fact that `PT` under BoTS failures surpasses its baseline, while this is not the case for `Rr`.

## 6.3 Batched-merge variant

The last protocol variant of Section 5 lets each node process the **entire** incoming queue $I_u$ in one round: one merge per distinct partition received, then a single local update. We compare batched-merge (BM) against the single-merge base (SM, i.e. the protocol of Section 6.1) for the three survivors of Section 6.1 in a complete graph topology.

Figure 6 shows the batched-merge trajectories alongside the single-merge baseline taken from Figure 2. At $\beta = 100$ (top row), BM and SM are visually indistinguishable for all three strategies. At $\beta = 0.1$ (bottom row), the two laggards of Section 6.1 both gain accuracy under BM: `PT` by 9.61% on HAR and 10.01% on MNIST, `Rr` by 7.37% on HAR and 3.51% on MNIST. `Ri` sits at the BM/SM common ceiling and gains essentially nothing. The ranking `Ri` $\geq$ `Rr` $\geq$ `PT` from Figure 2 is preserved under BM.

For `PT`, the gain tracks its **burstiness**: PTGL emits up to one proactive plus two reactive model messages per round (Section 5), so a receiver's incoming queue $I_u$ can hold several messages between rounds. Under SM the receiver processes only one per round, and the rest are deferred; the deferred merges are paid for as wasted training rounds on the corresponding partitions. Under BM the same receiver flushes all pending merges before its single local update, so PTGL's bursts translate into integration progress rather than accumulated delay. `Rr` emits exactly one model message per round, like `Ri`, so per-receiver arrivals are similarly distributed and burstiness alone does not explain its gain. The likely mechanism is different: `Rr` picks a partition uniformly at random, which under skewed local data leaves some partitions chronically under-served at any given receiver. BM lets the receiver catch up on these partitions whenever they do arrive together with others in the same round, which `Rr` already needs more than `Ri` does because `Ri` picks the local under-served partition directly. The gain measured here is therefore not about absorbing bursts but about recovering from random misalignment.

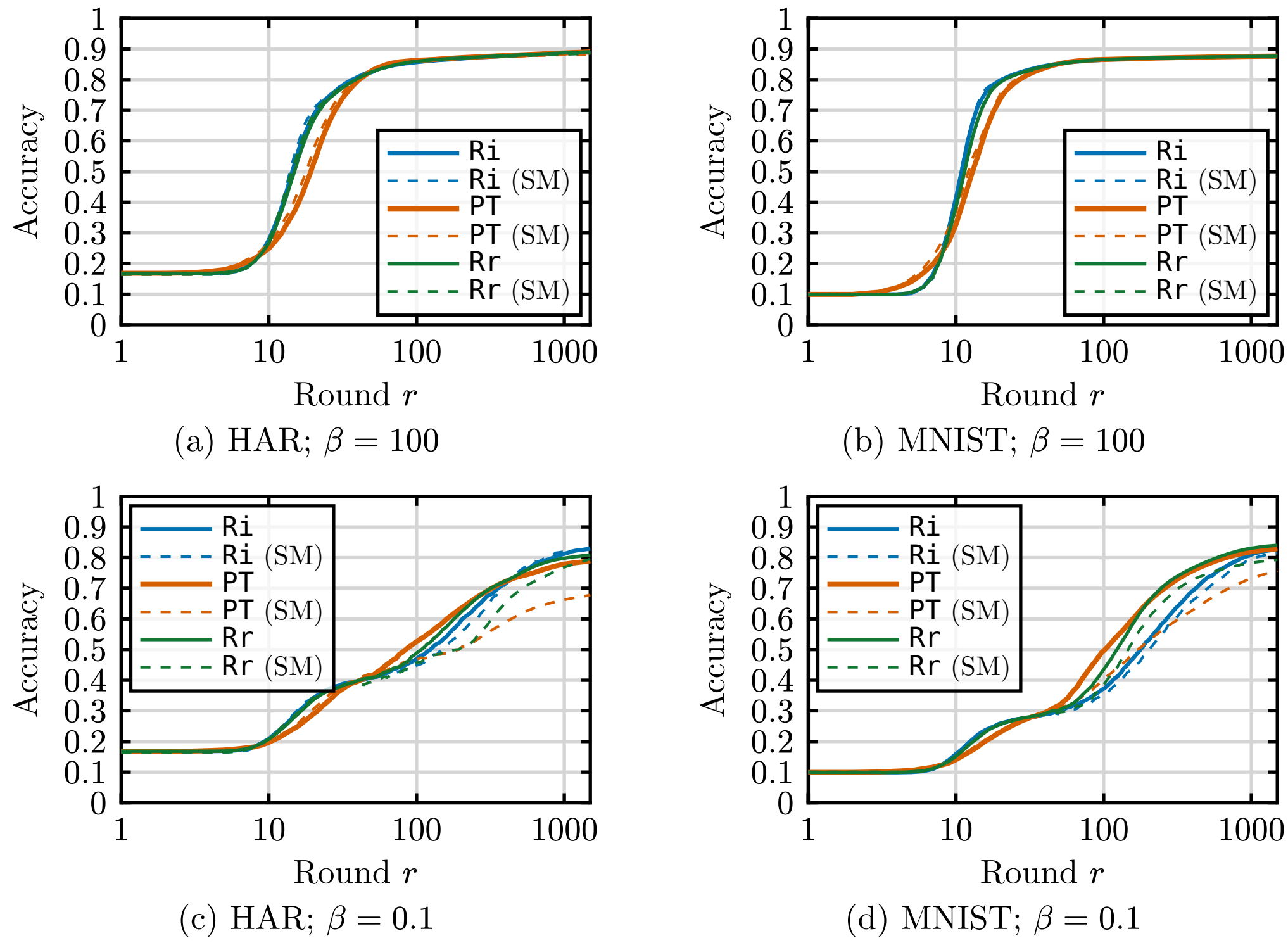

**Figure 7** Random 20-regular graph, batched-merge (solid) vs single-merge (dashed) for `Ri`, `Rr`, and `PT`. Rows: $\beta \in \{100, 0.1\}$. Columns: HAR, MNIST. At $\beta = 100$ (top row) BM and SM are visually indistinguishable for all three strategies. At $\beta = 0.1$ (bottom row), BM has essentially no effect on `Ri` (already near-saturation) but lifts both laggards.

Similarly to Figure 6, Figure 7 shows the batched-merge trajectories alongside the single-merge baseline, but in a random 20-regular graph. Similar to the observations made in Section 6.1 and Section 6.2, in the homogeneous case, results in the complete and random 20-regular graphs are close. In the heterogeneous case, we also observe that `PT` and `Rr` benefit from BM while `Ri` does not significantly gain from it.

## 7 Conclusion

We transposed peer-to-peer live-streaming strategies [5] to partitioned gossip learning and evaluated ten partition- and neighbor-selection strategies against the state-of-the-art token-based algorithm of Hegedüs et al. [10], by elimination across three settings: fault-free, adversarial 30% crash of best-performing nodes, and batched-merge. In a complete graph topology, one strategy emerges as the consistent leader: `Ri`, which sends the locally least-trained partition to a uniformly random neighbor. `Ri` matches or outperforms PTGL in every configuration tested, while requiring neither token counters nor metadata exchange; it reads only the local age vector $t_u$, and its margin over PT is largest under the realistic combination of adversarial crashes and data heterogeneity.
However, in a random 20-regular graph topology, Ri outperforms PT and Rr in the fault-free and batched-merge scenario, but is inefficient in the adversarial 30% crash of best-performing nodes when data-distribution is skewed.

In a complete graph topology, the structural reading is that PTGL's token machinery (counters, $\sigma$ and $\varphi$ functions, reactive bursts) can be replaced by the rule "send the least-trained partition you have", with no loss of accuracy in the homogeneous regime and substantial gains in the heterogeneous one. Comparing Ri to its degree-zero baseline Rr (random partition, random neighbor, no age) isolates the contribution of the local age vector: under heterogeneity, Ri leads Rr substantially, both under the fault-free protocol and under BoTS crashes. The local age information thus carries the load, whereas the token machinery built on top of it actually underperforms uniform random selection in this regime. The batched-merge variant adds a complementary observation: at $\beta = 0.1$ it lifts both PT and Rr while leaving Ri essentially unchanged. PT pays for its burstiness, Rr for random partition picks under heterogeneity, and BM mitigates both pathologies (see Section 6.3 for the detailed mechanisms); Ri exhibits neither and the merge policy is moot. The ranking Ri $\geq$ Rr $\geq$ PT is preserved under BM. From the perspective of distributed systems, partition scheduling under Ri reduces to a stateless, parameter-free local rule, which simplifies analysis, deployment, and reasoning about failure modes. In a random 20-regular graph topology, the observations made in the complete graph topology remain, except for one particular setting (adversarial failures and skewed data distribution). Prioritizing the less trained partition is then inefficient, when compared to the other 2 strategies selected, which use random partition selection.

This study has clear limitations. Our PTGL implementation transcribes the original protocol into the synchronous round-based model of Section 5 with the reactive burst capped at 2 messages per round; we did not independently validate this against the published asynchronous behavior, so all PTGL numbers should be read as relative to our simulator. Beyond these limitations, extending the comparison to deeper architectures, richer benchmarks, asynchronous settings [3, 4], and a formal convergence analysis of Ri, are all obvious next steps.

# A Background

## A.1 Supervised Machine Learning

Machine learning (ML) aims to produce, from data, an algorithm that solves a given task. Within supervised ML, we focus on the *classification problem*. Given a dataset $\mathcal{D} = \{X, Y\} = \left\{\left(x_1, ..., x_{n_\mathcal{D}}\right), \left(y_1, ..., y_{n_\mathcal{D}}\right)\right\}$, with $n_\mathcal{D} = |X| = |Y|$, $x_i \in \mathbb{R}^d$ and $y_i \in [\![0, K-1]\!]$, where $K \geq 2$ is the number of classes, the goal is to map an instance $x$ to its label $y$ using a *hypothesis function* $h_{w,b}$. Training minimizes the objective function Eq. (2) over the weights $w$ and bias $b$, using a *loss function* $\ell$ that measures the discrepancy between the prediction $h_{w,b}(x)$ and the true label $y$.

$$J(w, b) = \frac{1}{n_\mathcal{D}} \sum_{i=1}^{n_\mathcal{D}} \ell\big(h_{w,b}(x_i), y_i\big) \tag{2}$$

Logistic Regression is a binary classification algorithm where $y_i \in \{0, 1\}$ and $h_{w,b}$ is defined in Eq. (3)[4]. After fixing a threshold $m$, $x$ is classified according to whether $h_{w,b}(x) \geq m$.

$$h_{w,b}(x_i) = \frac{1}{1 + e^{-w^T x_i + b}} \tag{3}$$

The standard choice for $\ell$ in this setting is the regularized cross-entropy $\ell(x, y) = -\big(y \log\big(h_{w,b}(x)\big) + (1-y)\log\big(1 - h_{w,b}(x)\big)\big) + \frac{\lambda}{2}\,\|w\|_2^2$. The optimization is typically performed by Stochastic Gradient Descent [6], yielding the update rules Eq. (4) and Eq. (5).

$$w_{t+1} = w_t - \eta \nabla_w \ell\big(h_{w,b}(x_i), y_i\big) \tag{4}$$

$$b_{t+1} = b_t - \eta\big(h_{w,b}(x_i) - y_i\big) \tag{5}$$

In Eq. (4) and Eq. (5), $\eta$ and $\lambda$ are hyper-parameters (not adjusted by the optimization process), called the *learning rate* and *L2 regularization coefficient* respectively.

In this paper we use *multinomial* Logistic Regression, the natural generalization to $K > 2$ classes. Letting $\mathcal{M}_{d,K}$ denote the space of $d \times K$ matrices, the weights become $w \in \mathcal{M}_{d,K}$ (one logistic regression per class) and the prediction for an input $x$ is $y_{\text{pred}} = \text{argmax}\left(\text{softmax}\left(h_{w,b}(x)\right)\right)$.

## A.2 Federated Learning

In the standard supervised setting above, a single machine $s$ holds both the dataset $\mathcal{D}$ and the trained model $M$, which it serves to clients on request. Centralizing $\mathcal{D}$ on $s$ is problematic when the data is sensitive (e.g. medical records). Federated Learning (FL) addresses this by training a shared model across multiple devices without ever pooling their datasets, while aiming for accuracy comparable to the centralized case. The standard FL algorithm, *FederatedAveraging*, was introduced by McMahan et al. [14].

The server $s$ no longer owns a dataset; instead, it builds $M$ by aggregating local models $M_i$ trained by clients $c_1, ..., c_n$, each $c_i$ solving the local version of Eq. (2) on its own dataset $D_i$. Clients send their local models to $s$, which aggregates them and sends the result back for the next round. This procedure converges to the parameters $(w, b)$ that minimize the global objective Eq. (6).

[4] The name comes from the fact that this function is the standard logistic function $f(x) = \frac{L}{1+\exp(-k(x-x_0))}$ with $L = k = 1$ and $x_0 = 0$.

$$\frac{1}{n}\sum_{i=1}^{n} J_{i(w_i,b_i)} \tag{6}$$